\documentclass[prd,amsmath,amssymb,superscriptaddress,preprintnumbers,twocolumn,10pt]{revtex4-1}

\usepackage{graphicx}
\usepackage{dcolumn}
\usepackage{bm}
\usepackage{amssymb}
\usepackage{latexsym}
\usepackage{booktabs}
\usepackage{amsmath}
\usepackage{multirow}
\usepackage{url}

\usepackage{float}
\usepackage[colorlinks=true, linkcolor=red, citecolor=blue]{hyperref}

\usepackage[normalem]{ulem}
\usepackage{color}
\usepackage{array}
\usepackage{enumerate}
\usepackage{lineno}

\begin{document}

\title{
Constraining primordial black hole abundance with Insight-HXMT
}

\author{Chen Yang}
\affiliation{Liaoning Key Laboratory of Cosmology and Astrophysics, College of Sciences, Northeastern University, Shenyang 110819, China}

\author{Jun-Da Pan}
\affiliation{Liaoning Key Laboratory of Cosmology and Astrophysics, College of Sciences, Northeastern University, Shenyang 110819, China}

\author{Xin Zhang}\thanks{Corresponding author}
\email{zhangxin@neu.edu.cn}
\affiliation{Liaoning Key Laboratory of Cosmology and Astrophysics, College of Sciences, Northeastern University, Shenyang 110819, China}
\affiliation{National Frontiers Science Center for Industrial Intelligence and Systems Optimization, Northeastern University, Shenyang 110819, China}
\affiliation{MOE Key Laboratory of Data Analytics and Optimization for Smart Industry, Northeastern University, Shenyang 110819, China}

\begin{abstract}
Primordial black holes (PBHs), a major candidate for dark matter, have been extensively constrained across most mass ranges. However, PBHs in the mass range $10^{17}$--$10^{21}$ g   remain a viable explanation for all dark matter. In this study, we use observational data from the Hard X-ray Modulation Telescope (Insight-HXMT) to refine constraints on PBHs within the mass range $2\times10^{16}$--$5\times10^{17}$ g. Our analysis explores three scenarios:  directly using observational data, incorporating the astrophysical background model (ABM), and employing the power-law spectrum with an exponential cutoff. Our research results indicate that although Insight-HXMT does not have an advantage in the first two scenarios, when considering the power-law model, its exceptional sensitivity in the hard X-ray regime and sufficiently high upper energy limit significantly strengthen the constraints on PBHs with masses greater than $10^{17}$ g compared to previous limits. Furthermore, the exclusion limit for PBHs as dark matter has reached $4\times10^{17}$ g, which is comparable to the current threshold.
\end{abstract}

\maketitle

\section{Introduction}

Dark matter remains one of the most profound mysteries in modern astrophysics and cosmology. Observations such as Galactic rotation curves, gravitational lensing, and anisotropies in the cosmic microwave background provide compelling indirect evidence for its existence. However, despite these clues, the fundamental nature of dark matter continues to elude direct detection. Current models suggest that dark matter constitutes approximately 85\% of the total matter in the universe~\cite{WMAP:2010sfg,Planck:2015fie}, interacting with baryonic matter primarily through gravity and remaining invisible to electromagnetic observations. Over the past few decades, numerous candidates have been proposed, such as weakly interacting massive particles (WIMPs), sterile neutrinos, and axions. These candidates often require modifications to the Standard Model of particle physics, and none have been definitively detected to date.

Amid these ongoing searches, primordial black holes (PBHs) have emerged as an intriguing alternative, potentially bypassing the need for such modifications. PBHs are theoretical black holes that formed in the early universe due to the collapse of high-density fluctuations shortly after the Big Bang~\cite{Hawking:1971ei}. Unlike stellar black holes, PBHs can span a wide range of masses, from the Planck scale to billions of solar masses. This broad mass spectrum and early formation make PBHs relevant not only to the dark matter problem but also to various astrophysical phenomena. For instance, PBHs could account for the unexpected massive early galaxies observed by the James Webb Space Telescope (JWST)~\cite{Yuan:2023bvh,Liu:2022bvr}, and they offer a plausible explanation for the origin of the stellar-mass binary black holes detected by the LIGO and Virgo collaborations~\cite{Sasaki:2016jop,Ali-Haimoud:2017rtz}.

While many mass ranges for PBHs as the entirety of dark matter have been ruled out through observations like lensing, gravitational waves, and accretion, a relatively unconstrained mass window remains between $10^{17}$ g and $10^{21}$ g~\cite{Carr:2020gox,Carr:2021bzv}. This mass range is particularly notable because it circumvents many of the astrophysical and cosmological constraints that apply to PBHs at both lower and higher masses. Although several previous studies have attempted to impose observational constraints on the abundance of PBHs within this mass window \cite{Laha:2019ssq,Laha:2020ivk,Coogan:2020tuf,Iguaz:2021irx,Berteaud:2022tws}, \textit{\textbf{}}the results have been unsatisfactory, yielding only marginal improvements.  In this regime, Hawking radiation becomes a key mechanism for detecting PBHs and constraining their abundance. PBHs in this window are expected to emit radiation primarily in the keV--MeV energy range, making instruments sensitive to hard X-rays and soft $\gamma$-rays ideal for imposing tighter constraints on their abundance.

The first Chinese X-ray astronomy satellite, the Hard X-ray Modulation Telescope (Insight-HXMT) offers a broad energy range, large effective area, and high sensitivity, particularly in the hard X-ray band~\cite{Insight-HXMTTeam:2019dqg,Zhang:2019ikm}. Its sensitivity surpasses that of other instruments, such as the International Gamma-Ray Astrophysics Laboratory/Imager on Board the INTEGRAL Satellite (INTEGRAL/IBIS) and Rossi X-ray Timing Explorer/High-Energy X-ray Timing Experiment (RXTE/HEXTE)~\cite{Zhang:2019ikm}. These features make Insight-HXMT particularly well-suited for detecting Hawking radiation from PBHs in the unconstrained mass window, positioning it as an ideal tool for constraining PBH abundance within this range. 

In this study, we use diffuse X-ray background data to constrain the abundance of PBHs in the mass range of  $2\times10^{16}$ g to $5\times10^{17}$ g. Our analysis takes into account both the photon flux generated directly by Hawking radiation and the photons produced from the annihilation of positrons emitted during PBH evaporation. We include contributions to the diffuse X-ray background from both Galactic and extragalactic sources. The data used in our analysis are sourced from the diffuse X-ray background observed by Insight-HXMT~\cite{Huang:2022guk}. We evaluate three scenarios: directly using the observational data, incorporating an astrophysical background model (ABM), and applying the power-law model fitted from the data to extrapolate the upper energy limit. By leveraging Insight-HXMT’s advanced observational capabilities, our goal is to significantly improve  existing constraints on PBH abundance and advance the limits set by prior studies.

The structure of the paper is as follows: In Section~\ref{sec:2}, we detail the calculations of photon flux from PBH evaporation. Section~\ref{sec:3} presents the statistical analysis and constraints on PBH abundance using observational data and a power-law model from Insight-HXMT. Finally, Section~\ref{sec:4} summarizes our conclusions.

\section{The photon flux from PBH evaporation}\label{sec:2}

According to Hawking radiation theory, within the framework of the Standard Model, black holes emit all particles described by the Standard Model~\cite{Hawking:1974rv}. The particle emission spectrum resembles blackbody radiation, modulated by a greybody factor~\cite{Hawking:1975vcx}:
\begin{equation}
\frac{\partial^{2} N_{i}}{\partial E_{i} \partial t} = \frac{1}{2 \pi} \frac{\Gamma_{i}\left(E_{i}, M\right)}{e^{E_{i} / T_{H}}-(-1)^{2 s}}.  
\end{equation}
The variables are defined as follows: ${i}$ is the type of particle, ${s}$ is the spin of the particle, ${N}$ is the number of particles, ${E}$ is the energy, ${\Gamma_{i}}$ is the greybody factor, ${M}$ is the mass of the black hole,  and 
${T_{H}}$ is the Hawking temperature. The relationship between 
${M}$ and ${T_{H}}$ is given by:
\begin{equation}
T_{H}=\frac{M_{P}^{2}}{8 \pi M},  
\end{equation}
where $M_{P}$ is the Planck mass.

Among the particles produced by Hawking radiation, some can give rise to unstable particles, such as hadrons formed through the fragmentation of gluons and quarks~\cite{MacGibbon:1990zk,MacGibbon:1991tj}. These unstable particles undergo decay, producing stable secondary particles. In our work, we primarily focus on photons generated by Hawking radiation, including those directly emitted, secondary photons resulting from the decay of unstable particles produced by Hawking radiation, and photons produced via the annihilation of positrons emitted by Hawking radiation with electrons in the surrounding environment. {Photons from other processes, such as in-flight annihilation,
are not included in this study as their contribution to the total photon flux is negligible. Detailed calculations in Ref.~\cite{Tan:2024nbx} show that the photon flux yield from in-flight annihilation is approximately four orders of magnitude lower than the primary photon flux of Hawking radiation.} We use the code \texttt{BlackHawk} to compute the photon and positron spectra generated by Hawking radiation~\cite{Arbey:2019mbc}.

\subsection{Photons generated by Hawking radiation}

The photons produced by PBH evaporation consist mainly of two components: emissions from Galactic PBHs and those from extragalactic PBHs. In the following calculations, we assume a monochromatic mass distribution for the PBHs, with the rationale for this assumption explained in the next section.

The differential photon flux per unit solid angle from extragalactic PBHs is given by:
\begin{equation}
\frac{\mathrm{d} \phi_{\gamma}^{\text {ext }}}{\mathrm{d} E}=\frac{f_{\mathrm{PBH}} \Omega_{\mathrm{DM}} \rho_{c}}{4 \pi M} \int_{0}^{z_{\max }} \frac{\mathrm{d} z}{H(z)} \frac{\partial^{2} N_{\gamma}}{\partial E \partial t}(E(1+z)) ,
\end{equation}
where $f_{\mathrm{PBH}}$ denotes PBH abundance, representing the fraction of dark matter constituted by PBHs, and $N_{\gamma}$ is the total number of photons emitted per PBH. 
The Hubble parameter is
$H(z)=H_{0}\sqrt{\Omega_{\Lambda}+\Omega_{m}(1+z)^{3}}$,
with $H_{0}=67.36~\mathrm{km\,s^{-1}\,Mpc^{-1}}$ and $\Omega_{\Lambda}=0.6847$.
The matter density parameter is $\Omega_{m}=0.315$, which is the sum of
$\Omega_{\mathrm{DM}}=0.2645$ for dark matter and $\Omega_{b}\simeq0.0505$ for baryons.
The critical density is $\rho_{c}=9.1\times10^{-30}~\mathrm{g\,cm^{-3}}$~\cite{Planck:2018vyg}.

The maximum redshift is chosen as $z_\mathrm{max} \geq 10 $. Theoretically, $z_\mathrm{max}$ should correspond to the redshift at photon decoupling; however, this integral is not significantly sensitive to large values of $z_\mathrm{max}$.

The differential photon flux per unit solid angle from Galactic PBHs can be expressed as:
\begin{equation}
\frac{\mathrm{d} \phi_{\gamma}^{\text{gal}}}{\mathrm{d} E}=\frac{f_{\mathrm{PBH}} \bar{J}_{D}}{{4 \pi}M} \frac{\partial^{2} N_{\gamma}}{\partial E_{\gamma} \partial t} ,  
\end{equation}
where $\bar{J}_{D}$ is referred to as the astrophysical $J$-factor in some studies related to dark matter searches~\cite{Coogan:2021sjs,Boddy:2015efa,CTAO:2024wvb}, typically used to describe the dark matter density per solid angle along the line of sight. Its specific form is:
\begin{equation}\label{eq:J}
\bar{J}_{D} = \frac{1}{\Delta \Omega}
\int_{\Delta \Omega} \mathrm{d}\Omega
\int_{\text{l.o.s.}} \mathrm{d}l\, \rho_{\mathrm{DM}}(l,\psi).
\end{equation}
Here, $l$ is the distance along the line of sight, typically integrated to infinity, and $\psi$ is the angle between the line of sight and the direction of the Galactic center. In our analysis, we adopt a Navarro-Frenk-White (NFW) profile~\cite{Navarro:1996gj} for the dark matter distribution, parameterized as:
\begin{equation} \rho_{\mathrm{DM}}(r) = \frac{\rho_{0}}{\frac{r}{r_s}\left(1 + \frac{r}{r_s}\right)^2}, \end{equation}
where $\rho_{0}=0.839~\mathrm{GeV\,cm^{-3}}$ is the characteristic density, and $r_{s}=11~\mathrm{kpc}$ is the scale radius~\cite{deSalas:2019pee}.

It is important to note that, although the photon fluxes resulting from the evaporation of Galactic and extragalactic PBHs are theoretically expected to exhibit identical spectral features, a crucial distinction remains: ${\mathrm{d} \phi_{\gamma}^{\text{ext}}}/{\mathrm{d} E}$ represents the extragalactic photon flux integrated over redshifts from $z=0$ to $z=z_\mathrm{max}$, whereas ${\mathrm{d} \phi_{\gamma}^{\text{gal}}}/{\mathrm{d} E}$ corresponds solely to the Galactic photon flux at $z=0$. This difference leads to noticeably distinct spectral shapes, as shown in Fig.~\ref{fig:1}.

\subsection{Photons from positron annihilations}
When a positron and an electron interact, they can either annihilate directly, producing two 511 keV photons, or form a bound state known as positronium (Ps) before annihilation. If the spins of the electron and positron in the bound state are antiparallel, they form para-positronium, which annihilates into two photons with energies of 511 keV each. Conversely, if the spins are parallel, they form ortho-positronium, which annihilates into three photons with a broader energy distribution.  According to Ref.~\cite{Guessoum:2005cb}, when positrons produced by PBH evaporation interact with electrons in the cosmic medium, positronium formation ($f_\mathrm{P}=1$) is significantly more probable than direct annihilation ($f_\mathrm{P}=0$) under Milky Way-like conditions. Consequently, we consider only the scenario in which positronium is formed.
  
The photons from positron annihilation also consist of two components: the flux from extragalactic PBHs and the flux from Galactic PBHs, expressed as follows:
\begin{equation}\label{eq:ean}
\begin{aligned}
\frac{\mathrm{d} \phi_{\text {ann }}^{\text {ext }}}{\mathrm{d} E} = ~ & \frac{f_{\mathrm{PBH}} \Omega_{\mathrm{DM}} \rho_{c}}{4 \pi M} \frac{\mathrm{d} N_{e^{+}}}{\mathrm{d} t} \int_{0}^{\mathrm{z}_{\max }} \frac{\mathrm{d} z}{H(z)} \\
& \left(\frac{1}{2} \delta\left(E(1+z)-m_{e}\right)+\frac{9}{4} \frac{h_{3 \gamma}\left(\frac{E(1+z)}{m_{e}}\right)}{m_{e}}\right),
\end{aligned}
\end{equation}
\begin{equation}\label{eq:jan}
\begin{aligned}
\frac{\mathrm{d} \phi_{\text {ann }}^{\text {gal}}}{\mathrm{d} E} = \frac{f_{\mathrm{PBH}} \bar{J}_{D}}{4 \pi M} \frac{\mathrm{d} N_{e^{+}}}{\mathrm{d} t} 
 \left(\frac{1}{2} \delta\left(E-m_{e}\right)+\frac{9}{4} \frac{h_{3 \gamma}\left(E/{m_e}\right)}{m_{e}}\right).
\end{aligned}
\end{equation}
Here, $N_{e^{+}}$ represents the total number of positrons emitted by a single PBH, and $h_{3 \gamma}(E/{m_e})$ denotes the energy distribution spectrum of photons produced by ortho-positronium annihilation~\cite{Ore:1949te,Manohar:2003xv}. Its explicit form is given by
\begin{equation}
\begin{aligned}
h_{3 \gamma}(x) &= \frac{2}{\pi^{2}-9} \Bigg[ \frac{2-x}{x} + \frac{(1-x) x}{(2-x)^{2}} \\
&\quad - \left(\frac{2(1-x)^{2}}{(2-x)^{3}} - \frac{2(1-x)}{x^{2}} \right) \log(1-x) \Bigg],
\end{aligned}
\end{equation}
where $x=E/{m_e}$. Considering that the ratio of para-positronium to ortho-positronium generated by positrons is 1:3, and the ratio of the number of photons produced by para-positronium to ortho-positronium is 2:3, the factors $\frac{1}{2}$ and $\frac{9}{4}$ are derived for the above equations.

\begin{figure}[htbp] 
\centering\includegraphics[width=0.9\columnwidth]{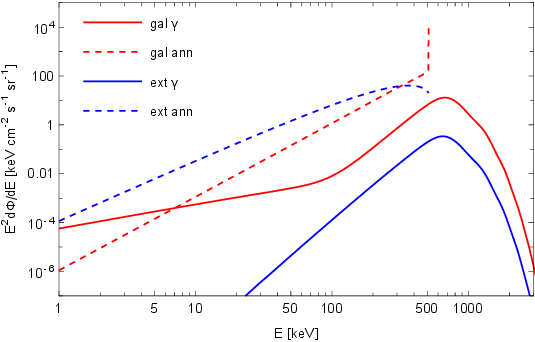} 
\caption{Various components of the total photon flux in Eq.~(\ref{eq:tf}) for $M=10^{17}$ g and $f_{\mathrm{PBH}}=1$. The red line represents the photon flux from Galactic sources, the blue line represents the photon flux from extragalactic sources. The solid line denotes the photon flux directly emitted by Hawking radiation, while the dashed line corresponds to the photon flux produced from positron annihilation.}
\label{fig:1}
\end{figure}

Finally, we obtained the total photon flux generated by the evaporation of PBHs:
\begin{equation}\label{eq:tf}
\psi^{\text{PBH}}=\frac{\mathrm{d} \phi^{\text{PBH}}}{\mathrm{d} E}=\frac{\mathrm{d} \phi_{\gamma}^{\text {ext }}}{\mathrm{d} E}+\frac{\mathrm{d} \phi_{\gamma}^{\text{gal}}}{\mathrm{d} E} +\frac{\mathrm{d} \phi_{\text {ann }}^{\text {ext }}}{\mathrm{d} E}+\frac{\mathrm{d} \phi_{\text {ann }}^{\text {gal}}}{\mathrm{d} E}. 
\end{equation}
In Fig.~\ref{fig:1}, we plot the photon flux from various components with $M=10^{17}$ g and $f_{\mathrm{PBH}}=1$.
From the figure, we observe that the photon flux directly emitted by Galactic PBHs exceeds that from extragalactic PBHs by more than an order of magnitude at the peak. Furthermore, for the total photon flux below 511~keV, the contribution from positron annihilation associated with Galactic and extragalactic PBHs dominates. In the Insight-HXMT energy band (1--250 keV), the sensitivity is dominated by positron\textendash electron annihilation, and the resulting 95\% CL limits are primarily set by the annihilation component.

\section{Constraints with Insight-HXMT data}\label{sec:3}
This section presents constraints on PBH abundance derived from diffuse X-ray background data collected by the Insight-HXMT satellite. Initially, the constraints are calculated using observational data without considering contributions from astrophysical backgrounds. To refine the analysis, the model incorporates an astrophysical background to account for contributions from sources such as star-forming galaxies (SFGs) and active galactic nuclei (AGNs), whose impact on the diffuse X-ray background significantly surpasses that of PBHs. By subtracting these contributions, the derived constraints become more accurate and realistic. Finally, since the observational data do not cover the entire energy range accessible to Insight-HXMT, we estimate the potential constraints that could be achieved with complete data coverage across the full energy spectrum.

The research in Ref.~\cite{Iguaz:2021irx} indicates that when considering the impact of positron annihilation on the abundance of PBHs, both the extended mass function and the spin of the PBHs lead to stronger constraints. When the mass distribution function is assumed to be log-normal, $f_{\mathrm{PBH}}$ exhibits negligible variation with increasing width parameter $\sigma$ until $\sigma$ exceeds a critical threshold; after exceeding the threshold, $f_{\mathrm{PBH}}$ starts to decay rapidly. When considering black hole spin, $f_{\mathrm{PBH}}$  exhibits a trend of continuously accelerating decay as the dimensionless spin parameter $\alpha$ increases. Therefore, in our work, to obtain more conservative and broadly applicable constraint results, we consider only the monochromatic, non-spinning PBHs, which represent the most general case.

\subsection{Constraints derived directly from observational data}

\begin{figure}[htbp] 
\centering\includegraphics[width=0.9\columnwidth]{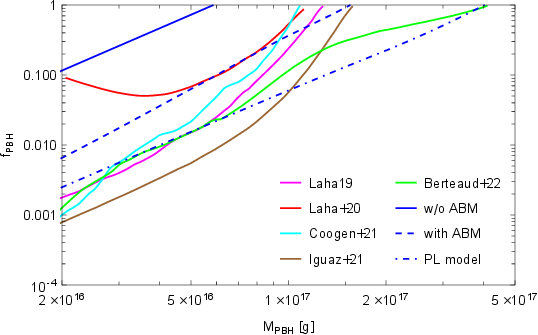} 
\caption{The constraint results in our work under three different scenarios: directly using observational data (blue solid line), incorporating the astrophysical background model (blue dashed line), and applying the power-law model fitted to the observational data (blue dotted-dashed line). These results are compared with existing constraints: magenta from~\cite{Laha:2019ssq}, red from~\cite{Laha:2020ivk}, cyan from~\cite{Coogan:2020tuf}, brown from~\cite{Iguaz:2021irx} and green from~\cite{Berteaud:2022tws}.}
\label{fig:2}
\end{figure}

To derive conservative constraints, we ensure that the total photon flux from PBHs does not exceed the observational data points measured by Insight-HXMT, including a 2$\sigma$ observational uncertainty, as described in refs.~\cite{Iguaz:2021irx,Coogan:2020tuf,Chen:2021ngo}. We utilized the measurements of the diffuse X-ray background flux and associated uncertainties presented in Figs.~8 and 9 in Ref.~\cite{Huang:2022guk}, which were primarily obtained from observations by the Low-Energy X-ray Telescope (LE) and the High-Energy X-ray Telescope (HE) onboard the Insight-HXMT. The observational data in Ref.~\cite{Huang:2022guk} were obtained from observation proposals with IDs P0101293, P0202041, and P0301293. We converted the sky region corresponding to each proposal from equatorial coordinates (R.A. and Dec.) to Galactic coordinates (l and b), and subsequently derived the range of $\psi$ for the $J$-factor in Eq.~(\ref{eq:J}) as $30^\circ$ to $145^\circ$ based on the converted coordinates. Our constraint results are shown in Fig.~\ref{fig:2}.

As shown in Fig.~\ref{fig:2}, when we directly use observational data to constrain the abundance of PBHs, our results are significantly weaker compared to previous studies~\cite{Laha:2019ssq,Laha:2020ivk,Coogan:2021sjs,Iguaz:2021irx,Berteaud:2022tws}. This discrepancy primarily arises from the limited availability of high-energy observational data in our analysis, as well as the absence of background subtraction for diffuse sources. These background sources contribute significantly to the measured flux, leading to a considerable underestimation of our constraints.

In Fig.~\ref{fig:2}, our result exhibits a simple linear relationship, which contrasts with the irregular shapes of the other curves. This is primarily due to two factors. First, with the exception of Iguaz et al. (2021), which adopts a similar approach to ours, all other studies in Fig.~\ref{fig:2} consider only the contribution from the Galactic photon flux component, ${\mathrm{d} \phi_{\gamma}^{\text{gal}}}/{\mathrm{d} E}$, in Eq.~(\ref{eq:tf}). Second, our analysis considers only the theoretical upper energy limit of Insight-HXMT, approximately 250 keV, while Iguaz's data are derived from multiple instruments with upper energy limits exceeding 10 MeV.

From Fig.~\ref{fig:2}, it is evident that within the energy range of Insight-HXMT, the contributions from ${\mathrm{d} \phi_{\text{ann}}^{\text{ext}}}/{\mathrm{d} E}$ and ${\mathrm{d} \phi_{\text{ann}}^{\text{gal}}}/{\mathrm{d} E}$ in Eq.~(\ref{eq:tf}) dominate the photon flux. In this regime, as $M_{\mathrm{PBH}}$ varies, the spectral shapes of ${\mathrm{d} \phi_{\text{ann}}^{\text{ext}}}/{\mathrm{d} E}$ and ${\mathrm{d} \phi_{\text{ann}}^{\text{gal}}}/{\mathrm{d} E}$ change mainly in terms of slope, resulting in minimal shifts in the optimal data point used to constrain $f_{\mathrm{PBH}}$, which consistently lies near the detector’s upper energy boundary. This leads to a nearly  linear relationship between $M_{\mathrm{PBH}}$ and $f_{\mathrm{PBH}}$ on a logarithmic scale.

However, when the maximum energy range exceeds 10 MeV, the situation changes. As $M_{\mathrm{PBH}}$ decreases and Hawking radiation intensifies, the ${\mathrm{d} \phi_{\gamma}^{\text{gal}}}/{\mathrm{d} E}$ component gradually becomes dominant. In this case, the optimal data points often correspond to the peak of ${\mathrm{d} \phi_{\gamma}^{\text{gal}}}/{\mathrm{d} E}$, the location and amplitude of which vary with $M_{\mathrm{PBH}}$. This results in more complex exclusion curves that no longer exhibit a simple linear behavior on a logarithmic scale.

\subsection{Situation with astrophysical background}

To mitigate the influence of diffuse background sources and thus obtain more accurate results, we adopted the ABM proposed in Ref.~\cite{Ballesteros:2019exr}. Since the dominant contributors to the diffuse X-ray background are blazars and AGNs, the model combines AGN and blazar emissions using a double power-law spectrum to fit the observational data. Its specific form is:
\begin{equation}\label{eq:ABM}
{ \psi^{\text{ABM}}}=\frac{\mathrm{d} \phi^{\text{ABM}}}{\mathrm{d} E} = \frac{A}{\left(E / E_b\right)^{n_1} + \left(E / E_b\right)^{n_2}},
\end{equation}
where the best-fit parameters are: $E_b=35.6966 ~\text{keV}$, $A= 0.0642~\text{keV}^{-1}~\text{s}^{-1}~\text{cm}^{-2}~\text{sr}^{-1}$, $n_1=1.4199$,  and $n_2=2.8956$. The observational data used to derive this fitting formula primarily originate from Solar Maximum Mission (SMM), Nagoya balloon, and High Energy Astronomy Observatory (HEAO).

After obtaining the astrophysical background model $\psi^{\mathrm{ABM}}$, we applied a ${\chi}^2$ fitting procedure to derive $f_{\mathrm{PBH}}$. The specific form of ${\chi}^2$ is
\begin{equation}
\chi^{2}=\sum\left[\frac{\psi^{\text {data }}-\psi^{\mathrm{PBH}}-\psi^{\mathrm{ABM}}}{\sigma^{\text {data }}}\right]^{2}.    
\end{equation} 
Here $\psi^{\text {data}}$ represents the differential photon flux per solid angle from diffuse X-ray background data, $\sigma^{\text {data}}$ is the associated uncertainty in  $\psi^{\text {data}}$.
We sum over all energy ranges of the data to calculate the ${\chi}^2$ value. To determine the 95\% confidence level bound on  $f_{\mathrm{PBH}}$, we calculate the ${\chi}^2$ value for each PBH mass M, requiring $\chi^{2}-\chi_{\min }^{2} \leq 3.84$. The constraints considering the astrophysical background model are shown in Fig.~\ref{fig:2}.

It is clear that when we remove the influence of the astrophysical background, the constraints on the PBH abundance are significantly improved. In the mass range $2\times10^{16}$--$5\times10^{17}$ g, the constraints are improved by approximately 1.5 orders of magnitude compared to the results obtained by directly using the observational data. While our results have been significantly improved, they only surpass some previous studies in the mass range above $10^{17}$ g. We attribute the weaker constraints to the lack of high-energy observational data. Furthermore, the method of using a double power-law spectrum to fit the observational data, which combines AGN and blazar emissions, to obtain the astrophysical background model, still appears somewhat rough and not sufficiently accurate. In future work, we plan to adopt the template analysis method from Ref.~\cite{Berteaud:2022tws} to analyze the Insight-HXMT observational data, in order to accurately determine the contribution of each background source to the diffuse X-ray background. This will further improve our constraint results.

\subsection{Extending the energy range of the diffuse X-ray background}
The observational data we used from Ref.~\cite{Huang:2022guk} spans an energy range of 1.5--120 keV, although the full observational range of Insight-HXMT extends from 1 to 250 keV. If data above 120 keV become available, Insight-HXMT could potentially provide more stringent constraint results. Considering this possibility, we performed a preliminary estimate of the constraints over the full 1--250 keV range.

Ref.~\cite{Huang:2022guk} used a power-law spectrum with an exponential cutoff to fit the observational data. Based on the best-fit parameters provided in the reference, we can model the diffuse X-ray background as follows:
\begin{equation}\label{eq:PL}
\frac{\mathrm{d} \phi_{\gamma}}{\mathrm{d} E}=9.57~E^{-1.402}~e^{-\frac{E}{55\mathrm{~keV}}}  \mathrm{~cm}^{-2} \mathrm{~s}^{-1} \mathrm{~keV}^{-1} \mathrm{~sr}^{-1}.   
\end{equation}

We assume that this power-law (PL) model remains valid up to the upper observational limit of Insight-HXMT's energy range. Therefore, within the 1--250 keV range, we require that the total photon flux from PBHs not exceed the photon flux of the diffuse X-ray background predicted by this model. The final constraint results also are presented in Fig.~\ref{fig:2}.

The constraint results, obtained by extrapolating the diffuse X-ray background model fitted to the Insight-HXMT observational data to the upper energy limit of the detector, are approximately 0.5 order of magnitude stronger than those based on the astrophysical background model. In this case, our observational results surpass those of most previous studies, providing stronger constraints than all others in the mass range above $10^{17}$ g. The exclusion limit for PBHs as dark matter has also been extended to $4\times10^{17}~\mathrm{g}$, which aligns with the best upper limit presented in Ref.~\cite{Berteaud:2022tws}. When comparing the ABM given by  Eq.~(\ref{eq:ABM}) with the PL model provided by Eq.~(\ref{eq:PL}), we found that they agree well within the range of observational data. However, when extrapolated beyond the observational range, the predictions of the PL model are significantly lower. This indicates that the PL
model may suffer from overfitting, which could lead to an overly optimistic interpretation of the final constraints. Due to the lack of sufficient observational data, we are unable to employ a ${\chi}^2$ estimation method to account for the influence of background diffuse sources. Therefore, we believe that with adequate observational data, Insight-HXMT will be capable of providing even more stringent constraint results.

\section{Conclusion}\label{sec:4}
In this work, we primarily utilize observational data from Insight-HXMT to improve the existing constraints on the abundance of PBHs in the mass range $2\times10^{16}$--$5\times10^{17}~\mathrm{g}$. We first consider the photon flux emitted by PBHs within this mass range, incorporating contributions from both Galactic and extragalactic sources. In this range, the photon flux generated by positron annihilation due to Hawking radiation cannot be neglected, particularly when the black hole temperature falls below 511 keV, where annihilation photons dominate. Therefore, we account for both the photon flux directly produced by Hawking radiation and the photons resulting from positron annihilation.

We then combine the diffuse X-ray background data measured by Insight-HXMT to derive constraints on PBH abundance under three different scenarios.

First, when we directly use the observational data from Insight-HXMT to derive the constraint results, due to the lack of high-energy observational data and the absence of background diffuse source subtraction, our results are weaker than those of previous studies. When we introduce the astrophysical background model, the constraints improve by approximately 1.5 orders of magnitude relative to the first scenario. Although the constraint results have improved significantly, we still do not have an advantage compared to previous studies.

Next, using the PL
model to estimate the constraints from Insight-HXMT with sufficient data above 120 keV, we find an improvement of approximately 0.5 orders of magnitude compared to the second scenario. After extending the energy upper limit, our observational results surpass all existing studies in the range above $10^{17}$ g. The upper limit of the constraints reaches $4\times10^{17}~\mathrm{g}$, which is consistent with the best current limits. 

In our analysis, the energy range of the data used spans from 1 to 120 keV, whereas the actual upper energy limit of Insight-HXMT observations extends up to 250 keV. Although we used the PL model to estimate the constraints based on data from 1 to 250 keV, the results remain less precise. Moreover, due to the lack of actual observational data, we are unable to remove the impact of diffuse background sources. This limitation prevents us from obtaining tighter constraints when using the power-law model. We need more observational data, especially in the 120--250 keV energy range, to improve our results. Furthermore, we applied an existing astrophysical background model without conducting a detailed analysis or disentangling individual sources within the data. Consequently, the results obtained are not the most accurate. We anticipate that by addressing these issues in future work, we will be able to achieve more accurate and robust results.

\begin{acknowledgments}
We thank Sai Wang, Yichao Li, and Chen Zhang for helpful discussions. We also thank Rui Huang for providing the observational data. This work was supported by the National SKA Program of China (Grants Nos. 2022SKA0110200 and 2022SKA0110203), the National Natural Science Foundation of China (Grants Nos. 12473001 and 12533001), the China Manned Space Program (Grant No. CMS-CSST-2025-A02), and the 111 Project (Grant No. B16009).
\end{acknowledgments}

\bibliography{main}

\end{document}